\documentclass[aps,prb,twocolumn,floatfix]{revtex4}

\usepackage{amsmath}		
\usepackage{amsfonts}
\usepackage{amssymb}
\usepackage{graphicx}		
\usepackage{bm}			

\usepackage{cancel}			
\usepackage{microtype}		

\begin{document}

\title{Kramers-Kronig, Bode, and the meaning of zero}

\author{John Bechhoefer}
\email[email: ]{johnb@sfu.ca}
\affiliation{Department of Physics, Simon Fraser University, Burnaby, B.C., V5A 1S6, Canada}

\date{\today}

\begin{abstract}
The implications of causality, as captured by the Kramers-Kronig relations between the real and imaginary parts of a linear response function, are familiar parts of the physics curriculum.  In 1937, Bode derived a similar relation between the magnitude (response gain) and phase.  Although the Kramers-Kronig relations are an equality, Bode's relation is effectively an inequality.  This perhaps-surprising difference is explained using elementary examples and ultimately traces back to delays in the flow of information within the system formed by the physical object and measurement apparatus.
\end{abstract}

\maketitle

\section{Introduction}

Dating from the dawn of modern quantum mechanics, the Kramers-Kronig (KK) relations,\cite{kronig26,kramers27} which connect the frequency-dependent real and imaginary parts of a linear response function, have wide application.  Part of their popularity resides in the generality of the KK relations:  they derive essentially from \textit{causality}---the response must follow the excitation and not precede it---and \textit{linearity}, the superposition of responses to different causes.\cite{toll56,sharnoff64}

Historically, the generality of the KK relations has made them a valuable tool, especially when measurements are limited or theory unclear.  For example, in \textit{particle physics} during the 1950s, the KK relations and associated \textit{sum rules} were helpful for making sense of scattering data and were an ingredient in the S-matrix theory used to analyze such experiments.\cite{nussenzveig72}  In \textit{optics}, Bode's gain-phase version of the KK relations has been widely used to analyze measurements of optical properties of materials, especially in reflection.  If the light source is incoherent, only the magnitude of the reflection coefficient (the optics version of the \textit{gain}) can be measured.  The Bode gain-phase relation then determines the phase.  Given the magnitude and phase, one can infer the index of refraction and absorption.\cite{jahoda57,smith85,peiponen09}

Although the KK relations are part of the ``standard lore" taught to physics students, the corresponding relation between the magnitude and phase of a complex response function is less well-known outside its specific applications to optics, even though it was derived by Bode in 1937\cite{bode37} and then popularized by him in an influential 1945 text.\cite{bode45}  Even less appreciated is that while the KK relations are an \textit{equality}, the Bode gain-phase relation is, in effect, an \textit{inequality}:  systems can have an ``extra" phase shift in their response that is greater than that given by the Bode relations.  This extra phase shift has been repeatedly rediscovered in various physics contexts\cite{toll56,grosse91,solli03} and is often remarked upon with surprise and explained in ways that are more complicated than they need to be.  

For reasons to be made clear, the Bode relation has been more appreciated by engineers than by physicists.  Drawing on the engineering literature, I derive and explain the Bode relation and give several simple examples where it is satisfied as an inequality rather than an equality.  Out of this exercise will come two insights:  first, a better understanding of the gain-phase relation itself and of its implications; second, a better appreciation that an experimental measurement reflects not only the dynamics of a physical system but also how excitations are made and how signals are received.  Choosing carefully the inputs and outputs to a physical system can help eliminate ``surprises."

\section{The Kramers-Kronig relations}
\label{sec:KK}

The Kramers-Kronig relations connect the real and imaginary parts of a causal linear response function, $G(t)$.  We interpret $G(t)$ as a Green function (or impulse-response function) that describes the response of a system at time $t$ after being excited by a delta function at time 0.  Causality implies that $G=0$ for $t<0$.  Linearity implies that the measurement $y(t)$ in response to an excitation $u(t)$ is given by
\begin{align}
	y(t) = \int_{-\infty}^\infty G(t-t')u(t') dt' \; \implies \; 
		y(\omega) = G(\omega) u(\omega)\,.
\label{eq:response}
\end{align}
In the language of engineers, $y(t)$ is the system \textit{output}, and $u(t)$ the system \textit{input}.  The expression for $y(\omega)$ follows from Fourier transforming and applying the convolution theorem and uses an ``overloading" notation where the same letter denotes a function in both its time- and frequency-domain representations.  

We are particularly interested in the frequency-domain response function,
\begin{align}
	G(\omega) \equiv G'(\omega) + iG''(\omega) =
	\int_{0}^\infty G(t) e^{i\omega t} \, dt \,,
\label{eq:Gfft}
\end{align}
where $G'$ and $G''$ are the real and imaginary parts of the response function, respectively.  Note that the lower limit in the integral in Eq.~\eqref{eq:Gfft} is 0, not $-\infty$, since $G(t)$ vanishes for $t<0$.

We can extend the Fourier transform into the complex-$\omega$ plane and consider the complex function $G$ over the $\omega$ plane.  Because $G(t)$ is causal, the integral in Eq.~\eqref{eq:Gfft} will converge for real $\omega$ if $G(t)$ vanishes fast enough at large $t$.  If it does, then the integral will converge even faster for $\omega$ in the upper half of the complex $\omega$-plane,\cite{stone09} and one can then show that $G(\omega)$ is \textit{analytic} in the upper-half plane.\cite{sharnoff64}  From these properties of $G$, it is straightforward to derive the Kramers-Kronig relations\cite{stone09, jackson99,greiner98} (cf. Appendix):
\begin{subequations}
\begin{align}
	G'(\omega) &=  \frac{2}{\pi} \, P \int_0^\infty 
	\frac{\omega' \, G''(\omega')}{\omega'^2 - \omega^2} \, d\omega' 
	\,, \qquad  \\[3pt]
	G''(\omega) &=  -\frac{2\omega}{\pi}  \, P \int_0^\infty 
	\frac{G'(\omega')}{\omega'^2 - \omega^2} \, d\omega' \,.
\end{align}
\label{eq:KK2}
\end{subequations}
In Eq.~\eqref{eq:KK2}, $P$ denotes the \textit{Cauchy principal value},\cite{stone09} which is defined by excluding  from the integration domain an infinitesimal region that is symmetrically distributed about the singular point, $\omega$ [see the Appendix and Eq.~\eqref{eq:kk2bode}, below].  Thus, for a causal response function $G(t)$, knowing the frequency dependence of the real part of the response function is equivalent to knowing its imaginary part, and vice versa.  Mathematically, the KK relations  in Eq.~\eqref{eq:KK2} are closely related to \textit{Hilbert transforms}.\cite{stone09}

\section{The Bode gain-phase relation}
\label{sec:bode-gainphase}

The Kramers-Kronig relations connecting the real and imaginary parts of a response function lead to an analogous connection between the amplitude and phase.  Let us assume that the response function $G(\omega)$ obeys the KK relations and that there are no values of $\omega$ in the upper-half of the complex $\omega$-plane for which $G(\omega) = 0$ (no \textit{zeros}).  If both conditions are met, we can
apply the Kramers-Kronig to the logarithm of the response function.\cite{which-plane}  We note that ln $G(\omega) = \ln \, |G(\omega)| + i \, \angle G(\omega)$, where $\angle G(\omega)$ is the phase of the complex number $G(\omega)$ at frequency $\omega$.  Then Eq.~\eqref{eq:KK2} gives
\begin{align}
	\angle G(\omega) = -\frac{2\omega}{\pi} \, P \int_0^\infty 
	\frac{\ln |G(\omega')|}{\omega'^2-\omega^2} \, d\omega' \,.
\label{eq:bode1}
\end{align}
 
As Bode recognized, Eq.~\eqref{eq:bode1} becomes more intuitive after integrating by parts.  First, we change variables: $\nu \equiv \ln \bigl( \tfrac{\omega'}{\omega} \bigr)$, or $\omega' = \omega \, e^\nu$, and $M(\nu) \equiv \ln |G(\omega')|$, giving
\begin{align}
	\angle G(\omega) &= -\frac{2}{\pi} \, P \int_{-\infty}^{\infty} \frac{\cancel{\omega} \, M(\nu)}{\cancel{\omega^2} \, \left( e^{2\nu}-1 \right)} \, \cancel{\omega} \, e^\nu \, d\nu \nonumber \\[3pt]
	&= -\frac{1}{\pi} \, P \int_{-\infty}^{\infty} \frac{M(\nu)}{\sinh \nu} \, d\nu \,,
\label{eq:bode2}
\end{align}
Because sinh is an odd function, only the odd part of $M(\nu)$ contributes in Eq.~\eqref{eq:bode2}.  Writing $M$ as the sum of odd and even functions, $M = M_o+M_e$, we have
\begin{align}
	&P \int_{-\infty}^{\infty} \frac{M(\nu)}{\sinh \nu} \, d\nu  & \nonumber \\[3pt]
	&= \lim_{\varepsilon \to 0^+}
	 \left\{ \int_{-\infty}^{-\varepsilon} + \int_{\varepsilon}^{\infty} \right\}
	  \left[ \frac{M_o(\nu) + M_e(\nu)}
		{\sinh \nu} \right] \, d\nu \nonumber \\[3pt]
	  &= 2 \lim_{\varepsilon \to 0^+}
	 \int_{\varepsilon}^{\infty}\frac{M_o(\nu)}{\sinh \nu} \, d\nu \,,
\label{eq:kk2bode}
\end{align}
since the even contributions $M_e$ cancel in the two integrals.  Now integrate by parts, noting that $\int \tfrac{d\nu}{\sinh \nu} = -\ln \, \coth \tfrac{\nu}{2}$.  Then, Eq.~\eqref{eq:kk2bode} becomes
\begin{align}
	2 \lim_{\varepsilon \to 0^+} \left\{ \int_{\varepsilon}^\infty \frac{dM_o}{d\nu} \ln \, \coth \frac{\nu}{2} \, d\nu -
	\left. M_o \ln \, \coth \frac{\nu}{2} \right|_{\varepsilon}^\infty \right\} \,.
\end{align}
The boundary terms vanish:  
\begin{enumerate}
\item As $\nu \to \infty$, ln coth $\tfrac{\nu}{2} \sim 2 e^{-\nu} = 2\tfrac{\omega}{\omega'}$.  As $\omega' \to \infty$, $|G(\omega')| \sim \omega'^{-n}$, since physical response functions vanish at infinite frequencies.  Then, $\ln |G| \sim -n \ln \omega'$.  Thus, $M_o(\nu) \ln \coth \tfrac{\nu}{2} \sim \tfrac{\ln \omega' }{\omega'} \to 0$.  

\item As $\nu \to 0$, ln~coth $\tfrac{\nu}{2} \approx -\ln \tfrac{\nu}{2}$.  Since $M_o$ is odd, $M_o \sim \nu + \mathcal{O}(\nu ^3)$.  Thus, $\nu \ln \nu \to 0$.  

\end{enumerate}

Thus, 
\begin{align}
	\angle G(\omega) &= -\frac{2}{\pi} \int_0^{\infty} \frac{dM_o}{d\nu} \, \ln \, \coth \frac{\nu}{2} \, d\nu \nonumber \\
	&= -\frac{1}{\pi} \int_{-\infty}^{\infty} \frac{dM}{d\nu} \, \ln \, \coth \frac{|\nu|}{2} \, d\nu \,,
\end{align}
where, in the last step, we used $2M_o = M(\nu)-M(-\nu)$.

\begin{figure}[tbh]
	\includegraphics[width=3.0in]{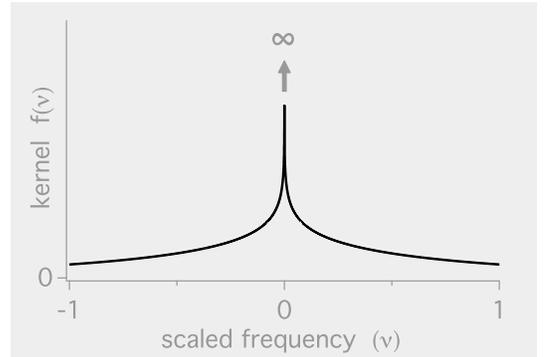}
	\caption{Kernel $f(\nu)$, with scaled frequency $\nu = \ln \tfrac{\omega'}{\omega}$.}
\label{fig:kernel}
\end{figure}

Finally, we have Bode's \textit{gain-phase relation}:\cite{sign,DCgain}
\begin{align}
	\angle \, G(\omega) &=  -\frac{\pi}{2}
	\int_{-\infty}^{\infty} \frac{dM}{d\nu} \, f(\nu) \, d\nu \nonumber \\
	f(\nu) &\equiv \frac{2}{\pi^2}  \ln \, \coth \frac{| \nu |}{2} \,, 
\label{eq:bode-gainphase}
\end{align}
where the kernel $f(\nu)$ resembles a broadened delta function (see Fig.~\ref{fig:kernel}) about $\nu=0$, where $\omega' = \omega$.  The kernel is normalized so that $\int_{-\infty}^\infty f(\nu) \, d\nu =1$.  

To understand the implications of the Bode relation intuitively, consider a frequency response $G(\omega) \sim \omega^{-n}$.  Such a relation typically holds at high frequencies for physical response functions.  For example, a low-pass filter has $n=1$, and a harmonic oscillator has $n=2$.  If this relation were to hold for \textit{all} frequencies $\omega > 0$, then
\begin{align}
	\frac{dM}{d\nu} = \frac{d \ln |G|}{d\ln \omega} = -n  \,,
\end{align}
and the phase delay is $\tfrac{\pi}{2} n$.  More generally, $n(\omega)$ is the \textit{local} value of  ln $| G(\omega)|$.  In that case, we note that the kernel $f(\nu)$ in Fig.~\ref{fig:kernel} resembles a broadened delta function, with most of its weight near $\nu=0$ ($\omega' = \omega$).  If $n(\omega)$ is constant over about a decade of frequency centered on $\omega$, then the Bode relation is, approximately,
\begin{align}
	\angle \, G(\omega) \approx -\frac{\pi}{2} \, 
	\frac{d \ln | G(\omega)|}{d \ln \omega}  \approx \frac{\pi}{2} \, n(\omega) \,.
\label{eq:bode-approx}
\end{align}
As a result, when the frequency response is graphed on \textit{Bode plots} with logarithmic frequency axes and a logarithmic magnitude axis, the phase lag is approximately the derivative of the magnitude curve times $\tfrac{\pi}{2}$.

\section{Bode relation and optical response}
\label{sec:optical-response}

In the Introduction, we noted that one of the main applications of the Bode gain-phase relation is in the determination of optical properties of materials.  The method is especially useful in the far infrared (IR), where there is a lack of bright, tunable, coherent sources.  Instead, one typically measures the \textit{reflectance} $R = |r^2|$ as a function of frequency $\omega$, where the \textit{reflection coefficient} $r$ is the complex linear response function $r = E_{\rm out}/E_{\rm in}$, the ratio of reflected to incident electric fields.  Because the source is incoherent, we cannot use the interference techniques that would normally help determine the phase.  Hence, only $R(\omega)$ is typically available and one must numerically integrate Bode's gain-phase relation, Eq.~\eqref{eq:bode-gainphase} with $R=G$, to determine the phase $\varphi(\omega)=\arg R(\omega)$.  For a thick sample whose reflectance is measured at normal incidence, we then use the Fresnel formula to write\cite{peiponen09}
\begin{align}
	r = \sqrt{R} \, e^{i\varphi} = \frac{n+ik-1}{n+ik+1} \,,
\label{eq:normal-incidence}
\end{align}
where $n(\omega)$ is the index of refraction and $k(\omega)$ the absorption coefficient.  Knowing the complex $r(\omega)$, we solve for $n$ and $k$ numerically.  

In practice, there are a number of issues, the most important of which is that the reflectance $R(\omega)$ can be measured only over a limited frequency range, and it is necessary to make some plausible guesses in order to extrapolate $R$ to all frequencies in the numerical integration of the Bode relation.\cite{grosse91}  As long as one works with normal incidence and thick samples, the method makes possible highly accurate inferences of the phase and hence of the material properties $n$ and $k$.  Indeed, as we have mentioned, the technique is the standard one for such measurements in the far-IR frequency range.  

For oblique incidence or thin films, there are many cases where the phase that is deduced from the Bode relation underestimates the actual phase shift and hence leads to incorrect inferences for the material parameters.\cite{grosse91,peiponen09}  The failure of a naive application of the Bode relation in this and other cases thus motivates a closer look at the underlying physics.

\section{Non-minimum-phase response functions}
\label{sec:NMP}

Bode's relation, Eq.~\eqref{eq:bode-gainphase}, is an \textit{equality} for the set of response functions that are analytic in the upper-half plane (and thus obey the KK relations) and, in addition, have no zeros in the upper-half plane.  Yet many physical response functions that are causal and obey the KK relations do have zeros in the upper-half plane.  Such response functions have extra phase delays:  the phase lag at a given frequency is greater than that predicted by the Bode relation.  Because these response functions are physical---indeed, we will give several examples---it makes sense to view the Bode relation as an \textit{inequality} over the set of physical response functions.

As an example, consider the response function for a \textit{delay} $\tau$, with output $y(t) = u(t-\tau)$.  Fourier transforming, we have
\begin{equation}
	y(\omega) = e^{i\omega\tau} u(\omega) \,,
\end{equation}
and we can identify the response function of the delay as $G_{\rm delay} = e^{i\omega\tau}$.  Since $G_{\rm delay}(\omega)$ has an essential singularity at $|\omega| \to \infty$, the logarithm is not analytic at infinity, and the Bode relation does not apply.  On the other hand, a delay is a physically possible, causal response function, and its real and imaginary components, $G'(\omega) = \cos \omega\tau$ and $G''(\omega) = \sin \omega \tau$ satisfy the KK relations, as may be verified by substitution into Eq.~\eqref{eq:KK2}.

We can calculate directly the magnitude and phase: $|G_{\rm delay}|=1$ and $\angle G_{\rm delay} = \omega \tau$.  By contrast, if there is no delay, then $G_0 = 1$, which has $|G_0|=1$ but $\angle G_0 = 0$.  Thus, we have two response functions, with equal magnitude response, but differing in the phase lag.  Applying Bode's relation to \textit{both} response functions predicts zero phase lag for both. (The exponent $n =0$.)  

We have seen that if a response function contains a delay, the phase lag will exceed that predicted by the Bode relation.  Since causality precludes a phase \textit{advance}, we conclude that the Bode relation gives a minimum phase lag: such \textit{minimum-phase} (MP) response functions have the smallest phase lag that is compatible with a given magnitude response.  \textit{Non-minimum-phase}  (NMP) response functions have a larger phase lag.

In addition to an exact delay, there are other NMP response functions that act as approximate delays.  Consider the family $G_n(\omega)$ of $n$th-order rational (Pad\'e) approximations to the unit delay $G_{\rm delay} = e^{i\omega}$.  The first- and second-order approximations are
\begin{align}
	G_1 &= \frac{1+\tfrac{1}{2}i\omega}{1-\tfrac{1}{2}i\omega} \,,
	\quad G_2 =  \frac{1+\tfrac{1}{2}i\omega-\tfrac{1}{12}\omega^2}
	{1-\tfrac{1}{2}i\omega-\tfrac{1}{12}\omega^2} \,,
\end{align}
The functions $G_n$ all have unit magnitude.  For example,
\begin{equation}
	|G_1(\omega)| = \left[ \left( \frac{1-\tfrac{1}{2}i\omega}{1+\tfrac{1}{2}i\omega} \right)
		\left( \frac{1+\tfrac{1}{2}i\omega}{1-\tfrac{1}{2}i\omega} \right)\right]^{1/2} = 1\,.
\label{eq:allpass1}
\end{equation}

\begin{figure}[h!]
	\includegraphics[width=3.5in]{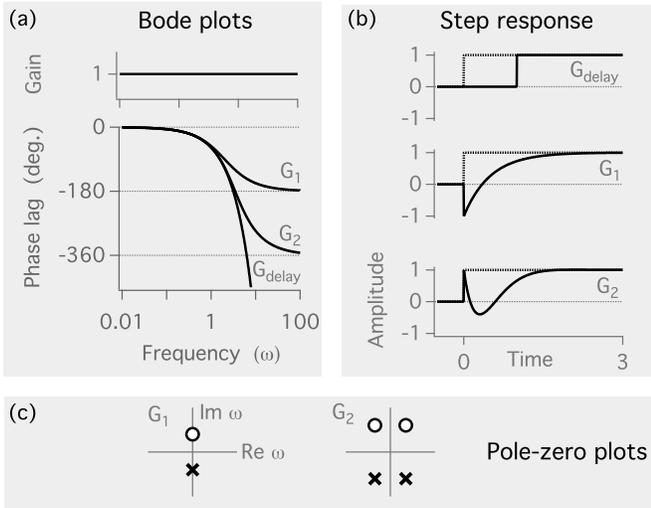}
	\caption{All-pass approximations to a unit delay.  (a)  Bode plots of $G_1$, $G_2$, and $G_{\rm delay}$.  (b)  Responses to a unit step. (c) Pole-zero plots for $G_1$ ($z=2i$, $p=-2i$) and $G_2$ ($z = 3i \pm \sqrt{3}$, $p = -3i \pm \sqrt{3}$).  Zeros are denoted by $\boldsymbol{\circ}$,  poles by $\boldsymbol{\times}$.}
\label{fig:pade}
\end{figure}

Response functions such as these with unit magnitude response at all frequencies are known as \textit{all-pass} functions.  (Because of this property, a Pad\'e expansion is more useful than a Taylor expansion.)  Here, we can easily verify that the high-frequency phase lag of $G_n$ is $n \tfrac{\pi}{2}$.  Figure~\ref{fig:pade}a shows the Bode plots and the responses to a unit step of $G_{\rm delay}$, $G_1$, and $G_2$.  We note that $G_1$ and $G_2$ do indeed approximate a delayed response.  Note that transients show ``inverse response" relative to the final value.  Indeed, the transient for $G_n(t)$ crosses the zero axis $n$ times before going to its asymptotic value of 1.  The other ``feature" of the Pad\'e approximants is that $G_1$ and $G_2$ have zeros in the upper half of the complex $\omega$ plane---as usual for non-minimum phase response functions---and poles in the ``mirror position" of the lower half of the $\omega$ plane (see Fig.~\ref{fig:pade}c).

Although the all-pass functions considered above might seem to be a special case of functions with zeros in the upper complex plane, an arbitrary NMP response function can always be decomposed into the product of a minimum-phase function (MP) times an all-pass function (AP).\cite{stable}   In symbols,
\begin{equation}
	G(\omega) = G_{\rm MP}(\omega) \, G_{\rm AP}(\omega) \,,
\label{eq:factorMP-AP}
\end{equation}
where $G_{\rm MP}$ is minimum phase and $G_{\rm AP}$ is all pass.

To see this result, we note that either $G(\omega)$ is already minimum phase or it has zeros at $\omega=\{iz_1\,, iz_2 \,, \ldots \}$.  Let us define the \textit{Blaschke product}\cite{toll56}
\begin{equation}
	G_{\rm AP}(\omega) = \left( \frac{z_1+i\omega}{z_1^*-i\omega} \right) \, 
	\left( \frac{z_2+i\omega}{z_2^*-i\omega} \right) + \cdots \,,
\label{eq:NMP-AP}
\end{equation}
which is all pass.  Note that since $G(t)$ is real, if there is a complex zero, its conjugate will also be a zero, as seen for $G_2(\omega)$ in Fig.~\ref{fig:pade}c.  

\begin{figure}[h!]
	\includegraphics[width=3.5in]{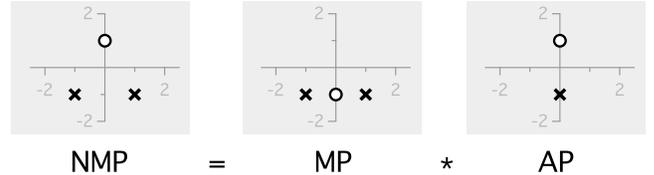}
	\caption{Decomposition of the non-minimum phase $G(\omega) = \tfrac{1+i\omega}{-\omega^2-2i\omega+2}$ into the product of a minimum-phase $\tfrac{1-i\omega}{-\omega^2-2i\omega+2}$ and an all-pass response function $\tfrac{1+i\omega}{1-i\omega}$.}
\label{fig:NMPdecomp}
\end{figure}

We then define $G_{\rm MP} = G / G_{\rm AP}$, which swaps the upper-plane zeros for their mirror reflection in the lower plane.  For example,
\begin{equation}
	G(\omega) = \underbrace{\frac{1+i\omega}{-\omega^2-2i\omega+2}}_
	{\text{non-minimum phase}} \, = 
	\underbrace{\left( \frac{1-i\omega}{-\omega^2-2i\omega+2} \right)}_
	{\text{minimum phase}} \,
	\underbrace{\left( \frac{1+i\omega}{1-i\omega} \right)}_{\text{all pass}} \,.
\label{eq:NMP-MP-AP}
\end{equation}
Note in Fig.~\ref{fig:NMPdecomp} how the zero of $G(\omega)$ at $\omega=i$ has been transferred to the all-pass function, while the minimum-phase function substitutes a ``reflected" zero at $\omega=-i$.  The poles at $-i \pm 1$ are  untouched.

\section{Examples and applications}
\label{sec:examples-applications}

We have seen that a response function has a phase lag that exceeds the amount predicted by Bode's gain-phase theorem when there is a zero in the upper half of the complex $\omega$ plane.  In this section, we present examples of systems that show such NMP behavior.

\subsection{Flexible and multimode objects}

\begin{figure}[h!]
	\includegraphics[width=2.5in]{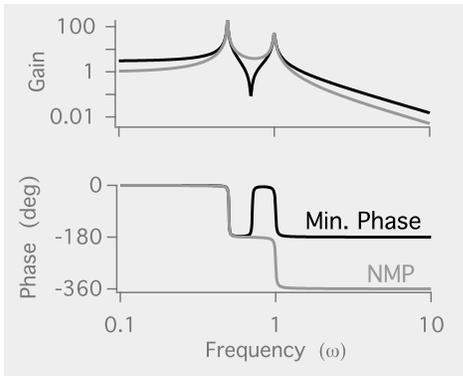}
	\caption{Two-mode dynamics Bode plot, from Eq.~\eqref{eq:TwoModeKKbode}, with $\alpha=1$, $\beta=2$, $\omega_0=0.5$.  Black line: minimum-phase case, $G_+$.  Gray line:  NMP case, $G_-$. A small amount of damping has been added to simplify the phase plots.}
\label{fig:TwoModeKKbode}
\end{figure}

One class of systems that are often NMP includes flexible objects---ones whose dynamics show contributions from many modes.  Often, the modes are extremely underdamped.  As a toy model of a flexible system, consider a system whose output adds contributions from two undamped modes, with frequencies scaled to 1 and $\omega_0$.  If the mode amplitudes are $\pm\alpha$ and $\beta$, the response in the frequency domain is
\begin{align}
	G_\pm(\omega) &= \pm \frac{\alpha}{1-\omega^2} +
	 \frac{\beta}{1-\frac{\omega^2}{\omega_0^2}} \nonumber \\[3pt]
	 &= \frac{\pm\alpha + \beta -\left(\frac{\pm\alpha}{\omega_0^2}+\beta \right) \omega^2}
	{(1-\omega^2) \, \left(1-\frac{\omega^2}{\omega_0^2} \right)} \nonumber \\[3pt]
	\implies \quad z^2 &=
	 \frac{\pm\alpha + \beta}{\frac{\pm\alpha}{\omega_0^2}+\beta} \,.
\label{eq:TwoModeKKbode}
\end{align}
Note how adding two oscillatory modes creates two zeros whose locations depend on the mode amplitudes $\alpha$ and $\beta$.  Figure~\ref{fig:TwoModeKKbode} shows Bode plots for a case where $\alpha$ and $\beta$ are chosen so that $G_+$ is minimum phase and $G_-$ is NMP.  The minimum-phase function has an asymptotic phase lag of $180^\circ$, as expected for a system of relative order = 2, while that of the NMP system is larger ($360^\circ$).  We have added a small amount of damping ($\zeta = 0.01$ for each mode) to soften the phase jumps and to keep responses finite.  The damping shifts the poles and zeros slightly below the real axis in the complex $\omega$-plane.  The poles then are strictly in the lower half of the $\omega$ plane, as they must be for a stable system.  With damping, the zeros give rise to finite-magnitude response minima, known as \textit{antiresonances}.  

\begin{figure}[tbh]
	\includegraphics[width=3.0in]{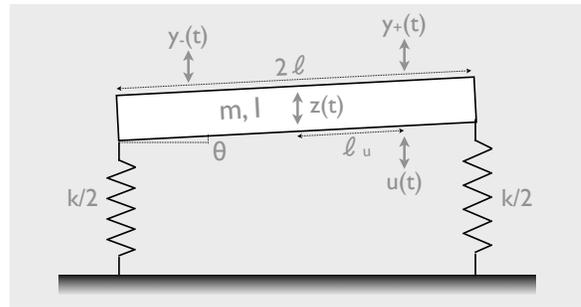}
	\caption{Object of mass $m$, length $2\ell$, and moment of inertia (about the center of mass) $I$ supported by two springs, each with force constant $\tfrac{1}{2}k$.  The vertical displacement from the center of mass is $z$. A force $u(t)$ is applied at right, at position $\ell_u$.  The force also acts as a torque.  The output is the vertical displacement, $y(x,t)$, measured at a point $x$ along the bar.  At the right edge ($x=\ell$), the measurement is denoted $y_+(t)$; at the left edge, $x=-\ell$, it is denoted $y_-(t)$.}
\label{fig:doyleNMP}
\end{figure}

Doyle et al.\cite{doyle92} have shown that this two-mode toy model has a simple realization, depicted in Fig.~\ref{fig:doyleNMP}.  A horizontal mass supported by two springs undergoes infinitesimal vertical displacement $z(t)$ and rotation $\theta(t)$.  If a force $u(t)$ is applied at a distance $\ell_u$ to the right of the center of mass, the equations of motion are
\begin{align}
	m\ddot{z} + kz &= u \nonumber \\
	I \ddot{\theta} + k \ell^2 \theta &= \ell_u u \,,
\label{eq:block-eqs}
\end{align}
where $m$ is the mass of the block, $I$ its moment of inertia about the center of mass, and $\tfrac{1}{2}k$ the spring constants.  Now---and this is the important step---consider measuring the position of the block at one of two places $y_{\pm} = z \pm \ell_u \theta$.  The first, $y_+(t)$, is located where the force is applied; the second, $y_-(t)$, is at the symmetric location,  $\ell_u$ to the left of the center of mass.  If we Fourier transform Eq.~\eqref{eq:block-eqs} and compute the response of $y_{\pm}(\omega)$ to the input $u(\omega)$, we find a function of the form of Eq.~\eqref{eq:TwoModeKKbode}.  Indeed, $m=2$, $k = I = \tfrac{1}{2}$, $\ell=1$, and $\ell_u=\tfrac{1}{\sqrt{2}}$ gives the example plotted in Fig.~\ref{fig:TwoModeKKbode}.  Thus, if we measure the position of the block where the force is applied, the response is MP.  Measured at the opposite side, the response is NMP.  In engineering jargon, the former case is a \textit{collocated} measurement and the latter a \textit{non-collocated} measurement.

The situation is similar, if more complicated when considering a flexible object whose motion is the sum of many modes.  Because each mode has a $180^\circ$ phase shift, the alternating pattern of poles and zeros seen in Fig.~\ref{fig:TwoModeKKbode} persists, with a zero between each resonance.  If the system is minimum phase, the maximum phase shift will continue to be $180^\circ$, no matter how many modes are relevant.  Such ideas are relevant and important in the analysis of \textit{atomic force microscopes}, which use a flexible cantilever to probe a surface.  The speed at which one can scan a surface can be limited by the response of the cantilever to forces created by the variable surface topography.  Certain combinations of inputs and outputs lead to MP response, while others lead to NMP response.\cite{rubio-sierra06}  A design without unnecessary NMP zeros allows higher scan rates.

\subsection{Optical systems}

Optical systems provide many examples of zeros (for example, destructive interference).  Here, we give two quick examples where the response is non-minimum phase.  The first, which we have already introduced in Sec.~\ref{sec:optical-response}, occurs in the analysis of reflectance spectra, where the goal is to infer, from measured reflectances, the complex phase shift as a function of light frequency.  The generalization of the Fresnel formula, Eq.~\eqref{eq:normal-incidence}, to an oblique angle of incidence $\theta$ gives, for TM radiation on a thick sample,
\begin{align}
	r_{TM}(\omega) = \frac{n^2 \cos \theta - \sqrt{n^2-\sin^2\theta}}
	{n^2 \cos \theta + \sqrt{n^2-\sin^2\theta}} \,,
\label{eq:fresnel}
\end{align}
where $n(\omega)$ is the complex index of refraction of the material under study (in air, for simplicity).  As Peiponen and Saarinen discuss\cite{peiponen09}, the response function $r_{TM}(\omega)$ can have a upper-plane zero for complex $\omega$ for some combinations of $n$ and $\theta$.  In such circumstances, there will be phase shifts beyond what the Bode relation predicts.  If one does not take into account the extra phase shifts, the absorption inferred will be incorrect.  The easiest fix is to choose conditions (for example, a thick sample at normal incidence) where the response is minimum phase.  Unfortunately, for a thin film, zeros associated with Fabry-Perot resonances are typical at all angles.\cite{grosse91}  If conditions leading to zeros cannot be avoided, independent measurements are needed to determine the phase.  Measuring the reflectances at different angles is one possibility.  

The second example, due to Solli et al.,\cite{solli03} occurred in a recent analysis of phase-sensitive measurements of microwaves propagating through a waveplate.  Although the focus of the work was to show that birefringence could lead to superluminal group velocities, the authors also noted that the phase shift showed an abrupt increase when the analyzer polarization angle was rotated past 45$^\circ$ with respect to the optical axis of the waveplate.  Indeed, a linearly polarized wave incident at 45$^\circ$ with respect to the optical axis and analyzed at an angle $\beta$ has an electric field
\begin{align}
	E(\omega) \propto e^{i\phi_{TM}} \, \left[\sin \beta e^{i\Delta \phi}  + \cos \beta \right] \,,
\label{eq:solli}
\end{align}
where $\Delta\phi(\omega) =  \Delta n(\omega) \, \omega d/c$, with $d$ the thickness of the waveplate and $c$ the speed of light.  Here, $\Delta n(\omega)$ is the frequency-dependent birefringence of the waveplate and $\phi_{TM}$ is the phase shift of the TM wave.  In Eq.~\ref{eq:solli}, $E(\omega)=0$ when, for integer $m$,
\begin{align}
	\Delta \phi^* = -i \ln | \cot \beta | + 2\pi \bigl(m+\tfrac{1}{2} \bigr) \,.
\label{eq:zero-condition}
\end{align}
As $\beta$ is varied about 45$^\circ$, $\cot \beta$ is larger or smaller than 1, so that Im $\Delta \phi^*$ is larger or smaller than 0.  Since $\Delta \phi \sim \omega \Delta \, n(\omega)$, and since $\Delta n$ is approximately real (and positive) for the conditions of the experiment, the zeros determined by $\omega^* \propto \Delta \phi^*$ change from the lower to the upper half plane for $\beta > 45^\circ$.\cite{imaginary-part}

\subsection{Implications for Feedback control}

Upper-plane zeros and NMP response functions are more familiar in engineering than in physics.  The reason is that the extra delays lead to problems when attempting to embed a NMP system inside a feedback loop.\cite{bechhoefer05,astrom08}  The basic ideas are simple:  In a feedback loop, the goal is typically to regulate or track a reference signal.  Any difference (or \textit{error}) is used to generate a correction signal.  But phase lags due to the time it takes signals to propagate from input to output
can make the control have the wrong correction.  In particular, if a sinusoidal signal lags by 180$^\circ$, the correction will be exactly in the wrong direction (positive feedback).  If the amplitude grows each loop, there will be a runaway oscillatory instability.  NMP systems exacerbate this problem by adding to the phase lag.  In addition, the inverse response of the transients (Fig.~\ref{fig:pade}b) also complicates the control problem.  NMP response thus limits the amount of feedback gain that can be applied.\cite{inversion}

To return to a mechanical example, a bicycle is an unstable system that is stabilized when moving fast enough.  Assuming it is, the transfer function from the steering angle of the front wheel to the tilt of the bike from the vertical has a zero in the lower-half of the complex $\omega$ plane.  But if the bike is steered from the rear (with the derailleur assembly on the front wheel), then the zero is in the upper half plane.  Such bicycles turn out to be practically unrideable.  In 1970, Jones, in an article that was recently reprinted in \textit{Physics Today},\cite{jones06} described attempts to create an unrideable bicycle, using an intuitive approach that was only partly successful (but very amusing).  {\AA}str\"om et al. explain how an understanding of bicycle dynamics can be used to make a truly unrideable bicycle or, more helpfully, an easier-to-ride bicycle suitable for disabled children.  The authors use models of bicycle dynamics to introduce, in a very accessible way, a number of ideas about control theory.\cite{astrom05}

\section{General implications}

\subsection{Input and output connections matter}

Starting from the Kramers-Kronig relation between the real and imaginary parts of a linear response function, we derived the analogous Bode relation between the magnitude and phase.  But unlike the KK relations, the Bode relation is most usefully viewed as an inequality:  if there are zeros in the upper part of the $\omega$ plane, there will be an extra phase lag (non-minimum-phase system).  In addition to producing the occasional ``surprise" in an experiment, we saw that non-minimum-phase systems are difficult to control.  Physically, these non-minimum-phase systems often correspond to situations where the input and output are separated in some way, so that there is a delay for the signal to get from the input to the output.

In the examples discussed in Sec.~\ref{sec:examples-applications}, a consistent theme was that the particular choice of measured variable could determine whether the response function is or is not minimum phase.  Where and what you measure matters.  By contrast, the resonance frequencies of a system are independent of the measurement details.  In our toy example of the two-mode system, the zeros were functions of the amplitudes $\alpha$ and $\beta$, but the poles that give the resonance frequencies were not.  That is, the zeros of linear response functions depend on the details of the excitation and the sensor, but the poles depend on the intrinsic dynamics.  The poles thus seem ``more fundamental" than the zeros.  Still, real experiments have sensors to make observations and, usually, actuators to create some kind of controlled excitation.  Experiments always mix intrinsic dynamics with experimental details of input and output connections, and the two aspects always need to be separated.  One practical lesson from the engineers is to be proactive and eliminate an upper-plane zero by rearranging sensors---choosing a different position to measure the block displacement, rotating a polarization analyzer angle,  or by doing more radical changes such as adding more sensors.  Because the root of the problem lies in the connections of signals between the outside world and the system under study, redesigning those connections can help.

\subsection{Beyond linear response}
Although you might think that zeros and their related issues are special features of linear response, they are more general.  It is true that notions of phase shifts are linked to linear systems, as they reflect a response to a sinusoidal inputs.  In a nonlinear system, a sine-wave input generates an infinite set of harmonic sine-wave outputs, each with its own phase shift that depends on the amplitude of the input.  Although there have been attempts to generalize the Kramers-Kronig (and Bode) relations to a nonlinear case, their usefulness is not clear.\cite{peiponen09}

On the other hand, the concept of a zero is not special to linear systems.  All that is required is that the output be zero for a class of input signals,  so that when you measure an output, you do not know which of the input signals was responsible.  For example, consider the time-domain version of an all-pass filter,
\begin{align}
	\dot{x}(t) = - x(t) + u(t) - \dot{u}(t) \,.
\label{eq:trivial}
\end{align}
Fourier transforming Eq.~\eqref{eq:trivial} gives the response function $G(\omega) = \tfrac{1+i\omega}{1-i\omega}$, which has a NMP zero at $\omega=i$.  The zero here means that an input of the form $u(t) = u_0 \, e^{t}$ does not affect the output, no matter what the value of $u_0$ (and even though the input is diverging with time).  This \textit{signal-blocking} property is another important feature of zeros.  Clearly, though, we would arrive at the same conclusion for the \textit{nonlinear} equation $\dot{x} = f(x) + u - \dot{u}$, with $f(x)$ a nonlinear function.  Thus, there are nonlinear equations with the same ``pathology" as linear equations.  A natural formulation of the nonlinear generalization of zeros is based on geometrical tools.\cite{isidori95}

From Eq.~\eqref{eq:trivial}, we see that the output tells nothing about the amplitude, or even presence, of the input, $u_0$.  Such loss of information is a familiar idea from communication theory, where the equivalent statement is that the \textit{mutual information} between input and output is zero:  the output gives no information about a set of inputs.  The information-theory analysis of dynamical response is particularly attractive in that both nonlinear and stochastic effects can be accommodated in a natural way.\cite{mackay03}

\section*{Appendix}
We give here a brief derivation of the Kramers-Kronig relations.\cite{jackson99,stone09,greiner98}  If $G(t)$ is a causal response function and if the response to an impulse dies away quickly enough, then we can assume that $G(t) \to 0$ as $t \to \infty$ fast enough that the integral $\int_0^\infty G(t) e^{i\omega t} \, dt$ converges.  If so, then the integral converges even faster for complex $\omega$ with Im $\omega > 0$.  As a result, $G(\omega)$ is \textit{analytic} in the complex $\omega$-plane for all Im $\omega > 0$.  For simplicity, we will also assume that $G(\omega)$ has no poles on the real axis.

\begin{figure}[h!]
	\includegraphics[width=3.0in]{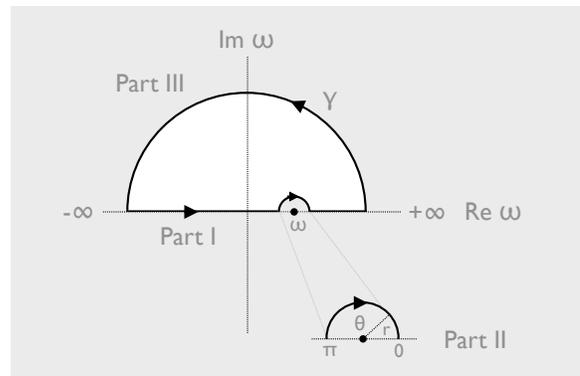}
	\caption{Path $\gamma$ for contour integral in Eq.~\eqref{eq:cauchy}.}
\label{fig:freq-KKcontour}
\end{figure}

Since $G(\omega)$ is analytic for Im $\omega \ge 0$, we can use \textit{Cauchy's Integral Theorem} to write
\begin{align}
	\oint_\gamma \frac{G(\omega')}{\omega'-\omega} \, d\omega' = 0 \,,
\label{eq:cauchy}
\end{align}
where the closed contour $\gamma$ is depicted in Figure~\ref{fig:freq-KKcontour}.  The semicircular indentation around the point $\omega$ is necessary since there is a pole in the integrand in Eq.~\eqref{eq:cauchy}.  

The integral in Eq.~\eqref{eq:cauchy} is divided into three parts, as labeled in Figure~\ref{fig:freq-KKcontour}. 
\begin{align}
	\text{Part I} = P\int_{-\infty}^{\infty} \frac{G(\omega')}{\omega'-\omega} \, d\omega' \,,
\end{align}
where $P$ denotes the \textit{principal value}, which simply is a notation to remind us that we have excluded an infinitesimal, symmetric region from the domain of integration.

Part II of the integral is a semicircle of radius $r \to 0$.  Writing $\omega' = \omega + re^{i\theta}$, we can approximate $G(\omega')$ by $G(\omega)$ and pull it out of the integral, leaving
\begin{align}
	\text{Part II} = G(\omega) \int_\pi^0 \frac{ir e^{i\theta}}{r e^{i\theta}} \, d\theta = -i \pi \, G(\omega) \,.
\end{align}

Finally, we assume that $G(\omega') \to 0$ fast enough for $|\omega'| \to \infty$ that  Part III $\to 0$ as the contour radius $R \to \infty$.  Then, from the Cauchy theorem, Parts I + II + III = 0, implying
\begin{align}
	G(\omega) = + \frac{1}{i\pi} P \int_{-\infty}^{\infty} \frac{G(\omega')}{\omega'-\omega} \, d\omega' \,.
\label{eq:freq-KKcomplex}
\end{align}
Writing $G=G'+iG''$ and isolating real and imaginary parts gives Eq.~\eqref{eq:KK2}.

\begin{acknowledgments}
I am grateful for funding from NSERC (Canada) and from Simon Fraser University (for sabbatical leave).  I thank Paul Martin, Karl {\AA}str\"om, Mike Plischke, Chris Homes, and Steve Dodge for their helpful suggestions.  I thank Suckjoon Jun and Bodo Stern for making my sabbatical stay at the Harvard FAS Center for Systems Biology a pleasant and fruitful one.

\end{acknowledgments}


\begin{thebibliography}{20}

\bibitem{kronig26}  R.~de L.~Kronig, ``On the theory of the dispersion of X-rays," J.~Opt.~Soc.~Am. \textbf{12}, 547--557 (1926).

\bibitem{kramers27} H.~A.~Kramers, ``La diffusion de la lumi\`ere par les atomes," Atti Cong. Intern. Fisica, (Transactions of Volta Centenary Congress) Como \textbf{2}, 545--557 (1927).

\bibitem{toll56}  J.~S.~Toll, ``Causality and the dispersion relation:  Logical foundations," Phys.~Rev. \textbf{104},  1760--1770 (1956).

\bibitem{sharnoff64} M.~Sharnoff, ``Validity conditions for the Kramers-Kronig relations," Am.~J.~Phys. \textbf{32}, 40--44 (1964).

\bibitem{nussenzveig72} H.~M.~Nussenzveig, \textit{Causality and Dispersion Relations}, (Academic Press, New York, 1972).

\bibitem{jahoda57} F.~C.~Jahoda, ``Fundamental absorption of barium oxide from its reflectivity spectrum," Phys.~Rev.~\textbf{107}, 1261--1265 (1957).

\bibitem{smith85}  D.~Y.~Smith, ``Dispersion theory, sum rules, and their application to the analysis of optical data," in \textit{Handbook of Optical Constants of Solids}, Vol.~2, ed. E.~D.~Palik (Academic Press, Orlando, 1985) 35--68.

\bibitem{peiponen09}  K.-E.~Peiponen and J.~J.~Saarinen, ``Generalized Kramers-Kronig relations in nonlinear optical- and THz-spectroscopy," Rep.~Prog.~Phys. \textbf{72}, 056401, 19 pp. (2009).

\bibitem{bode37}  H.~W.~Bode, US Patent 2,123,178.  Filed June 23, 1937.  Cf. H.~W.~Bode, ``Relations between attenuation and phase in feedback amplifier design," Bell Sys.~Tech.~ J. \textbf{19}, 421--454 (1940). 

\bibitem{bode45}  H.~W.~Bode, \textit{Network Analysis and Feedback Amplifier Design} (D.~van Nostrand and Co., New York, NY, 1945).


\bibitem{grosse91} P.~Grosse and V.~Offermann, ``Analysis of reflectance data using the Kramers-Kronig relations," Appl.~Phys.~A \textbf{52}, 138--144 (1991).

\bibitem{solli03}  D.~R.~Solli, C.~F.~McCormick, C.~Ropers, J.~J.~Morehead, R.~Y.~Chiao, and J.~M.~Hickmann, ``Demonstration of superluminal effects in an absorptionless, nonreflective system," Phys.~Rev.~Lett. \textbf{91}, 143906, 4 pp. (2003).

\bibitem{stone09}  M.~Stone and P.~Goldbart, \textit{Mathematics for Physics: A Guided Tour for Graduate Students} (Cambridge Univ. Press, Cambridge, UK, 2009).

\bibitem{jackson99}   J.~D.~Jackson,  \textit{Classical Electrodynamics}, 3rd ed. (John Wiley \& Sons, Inc., New York, NY, 1999).

\bibitem{greiner98}  W.~Greiner, \textit{Classical Electrodynamics}, (Springer, New York, 1998).

\bibitem{which-plane}  The engineering literature uses the complex $s$ plane and Laplace transform instead of the complex $\omega$ plane and Fourier transform.  Because the two are rotated by 90$^\circ$, our discussion of the analyticity properties in the lower and upper $\omega$ planes is equivalent to discussions in the engineering literature of analyticity properties of the left-hand and right-hand $s$ planes.

\bibitem{sign}  In the engineering literature, Eq.~\eqref{eq:bode-gainphase} usually has the opposite sign.  This difference traces back to the engineers' use of $e^{-i\omega t}$ rather than $e^{+i\omega t}$ in the forward Fourier transform.

\bibitem{DCgain}  In formulating the gain-phase relation, we assume that the DC gain (i.e., at $\omega=0$) is positive.  A negative DC gain can be regarded as an overall conversion factor between input and output rather than as an extra 180$^\circ$ phase shift.  Thus, for our purposes, both $G(\omega) = \tfrac{-1}{1-i\omega}$ and $\tfrac{1}{1-i\omega}$ have the same phase response.

\bibitem{stable}   We also need to assume that $G(\omega)$ has no poles in the upper half of the complex $\omega$-plane.  Such poles correspond to unstable, exponentially growing motion and also add to the phase delay of the response.  Only active systems, with external energy injection, can have such poles.

\bibitem{doyle92} J.~C.~Doyle, B.~A.~Francis, and A.~R.~Tannenbaum, \href{http://www.control.utoronto.ca/people/profs/francis/dft.html}{\textit{Feedback Control Theory}}, (Macmillan Publishing Co., New York, NY, 1992).

\bibitem{rubio-sierra06}  F.~J.~Rubio-Sierra, R.~V\'azquez, and R.~W.~Stark, ``Transfer function analysis of the micro cantilever used in atomic force microscopy," IEEE Trans.~Nanotech. \textbf{5}, 692--700 (2006).  The ``bad" NMP input-output combination is to apply a distributed-force and to measure the slope at the tip.  Such a situation occurs when the cantilever is excited by an external electric or magnetic field and the measurement uses the standard beam-displacement technique.

\bibitem{imaginary-part}  The small imaginary part of $\Delta n$ will shift the crossover value slightly from 45$^\circ$.

\bibitem{bechhoefer05}  J.~Bechhoefer, ``Feedback for physicists:  a tutorial essay on control," Rev.~Mod.~Phys. \textbf{77}, 783--836 (2005).

\bibitem{astrom08}  K.~J.~{\AA}str\"om and R.~M.~Murray, \textit{Feedback Systems}, (Princeton Univ. Press, Princeton, NJ, 2008).

\bibitem{inversion}  Another difficulty of NMP systems is that most control loops amount to approximately inverting the transfer function of a system.  Upon inversion, an upper-plane zero becomes an upper plane pole, and the inverse is unstable.

\bibitem{jones06}  D.~E.~H.~Jones, ``The stability of the bicycle," Phys.~Today \textbf{59}(9), 51--56 (2006).  Reprinted from Phys.~Today \textbf{23}(4) 34--40 (1970).

\bibitem{astrom05}  K.~J.~{\AA}str\"om, R.~E.~Klein, and A.~Lennartsson, ``Bicycle dynamics and control," IEEE Cont.~Syst.~Mag. \textbf{25}(4), 26--47 (2005).

\bibitem{isidori95}  A.~Isidori, \textit{Nonlinear Control Systems}, 3rd ed. (Springer, Berlin, 1995).

\bibitem{mackay03}  D.~J.~C.~MacKay, \textit{Information Theory, Inference, and Learning Algorithms}, (Cambridge Univ. Press, Cambridge, 2003).



\end{thebibliography}
\end{document}